\documentclass[journal=apchd5,manuscript=article,layout=twocolumn,articletitle]{achemso}
\usepackage{amsfonts}
\usepackage{amssymb}
\usepackage{amsmath}
\usepackage{graphicx}
\usepackage{bookmark}
\author{Francesco De Nicola}
\email{francesco.denicola@iit.it}
\affiliation{Graphene Labs, Istituto Italiano di Tecnologia, Via Morego 30, 16163 Genova, Italy}
\author{Nikhil Santh Puthiya Purayil}
\affiliation{Graphene Labs, Istituto Italiano di Tecnologia, Via Morego 30, 16163 Genova, Italy}
\affiliation{Physics Department, Universit\'a degli studi di Genova, Via Dodecaneso 33, 16146 Genova, Italy}
\author{Davide Spirito}
\affiliation{Nanochemistry Department, Istituto Italiano di Tecnologia, Via Morego 30, 16163 Genova, Italy}
\author{Mario Miscuglio}
\affiliation{Nanochemistry Department, Istituto Italiano di Tecnologia, Via Morego 30, 16163 Genova, Italy}
\affiliation{Chemistry and Industrial Chemistry Department, Universit\'a degli studi di Genova, Via Dodecaneso 33, 16146 Genova, Italy}
\author{Francesco Tantussi}
\affiliation{Plasmon Nanotechnologies, Istituto Italiano di Tecnologia, Via Morego 30, 16163 Genova, Italy}
\author{Andrea Tomadin}
\affiliation{Graphene Labs, Istituto Italiano di Tecnologia, Via Morego 30, 16163 Genova, Italy}
\author{Francesco De Angelis}
\affiliation{Plasmon Nanotechnologies, Istituto Italiano di Tecnologia, Via Morego 30, 16163 Genova, Italy}
\author{Marco Polini}
\affiliation{Graphene Labs, Istituto Italiano di Tecnologia, Via Morego 30, 16163 Genova, Italy}
\author{Roman Krahne}
\affiliation{Nanochemistry Department, Istituto Italiano di Tecnologia, Via Morego 30, 16163 Genova, Italy}
\author{Vittorio Pellegrini}
\affiliation{Graphene Labs, Istituto Italiano di Tecnologia, Via Morego 30, 16163 Genova, Italy}
\title{Multiband Plasmonic Sierpinski Carpet Fractal Antennas}
\begin{document}
\begin{tocentry}
\includegraphics[width=8cm,keepaspectratio]{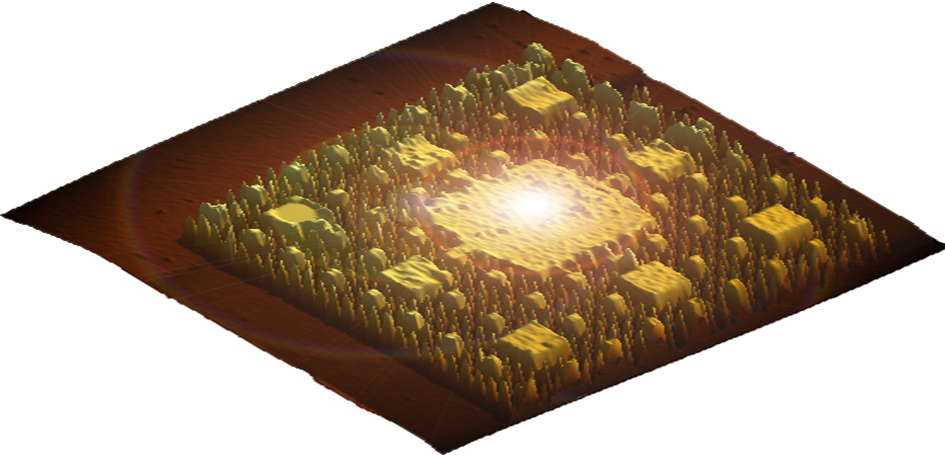}
Artistic illustration of an Au Sierpinski carpet.
\end{tocentry}
\begin{abstract}
Deterministic fractal antennas are employed to realize multimodal plasmonic devices. Such structures show strongly enhanced localized electromagnetic fields typically in the infrared range with a hierarchical spatial distribution. Realization of engineered fractal antennas operating in the optical regime would enable nanoplasmonic platforms for applications, such as energy harvesting, light sensing, and bio/chemical detection. Here, we introduce a novel plasmonic multiband metamaterial based on the Sierpinski carpet (SC) space-filling fractal, having a tunable and polarization-independent optical response, which exhibits multiple resonances from the visible to mid-infrared range. We investigate gold SCs fabricated by electron-beam lithography on CaF$_{2}$ and Si/SiO$_{2}$ substrates. Furthermore, we demonstrate that such resonances originate from diffraction-mediated localized surface plasmons, which can be tailored in deterministic fashion by tuning the shape, size, and position of the fractal elements. Moreover, our findings illustrate that SCs with high order of complexity present a strong and hierarchically distributed electromagnetic near-field of the plasmonic modes. Therefore, engineered plasmonic SCs provide an efficient strategy for the realization of compact active devices with a strong and broadband spectral response in the visible/mid-infrared range. We take advantage of such a technology by carrying out surface enhanced Raman spectroscopy (SERS) on Brilliant Cresyl Blue molecules deposited onto plasmonic SCs. We achieve a broadband SERS enhancement factor up to $10^{4}$, thereby providing a proof-of-concept application for chemical diagnostics. 
\end{abstract}
\indent Over the past decade, there has been a strong and ongoing interest in developing the design of fractal metamaterials for compact, multiband, and high-gain antennas intended for radio-frequency applications \cite{Werner2003}, THz resonators \cite{Miyamaru2008,Maraghechi2010,Agrawal2009} and lasers \cite{Mahler2010}, and microwave devices \cite{Wen2005,Werner2003}. Also, in the mid-infrared range plasmonic fractal structures have been reported to act as frequency-selective photonic quasi-crystals \cite{Negro2012,Hou2010,Huang2010,Aslan2016,Gottheim2015,Bao2007,Bao2008,Aygar2016,Aouani2013}. Besides a few attempts by Dal Negro et al. \cite{Gopinath2008,Lee2010a}, research in the visible range is still in an early stage. This is due to the expensive and challenging fabrication of nanofractals, resulting from the need of high-resolution electron- or ion-beam techniques.\\
\indent In this regard, plasmonic deterministic fractals \cite{Negro2012} offer the effective control and reproducibility of their optical properties, by adjusting their size, shape, and position. The scale-invariance property of deterministic fractals may be exploited to realize devices with a tailorable, self-similar, and multimodal spectral response. Furthermore, such fractals can provide a potential platform to study enhanced light-matter interactions. Moreover, the tunable plasmonic near-field enhancement may lead to the realization of advanced optoelectronic devices, such as solar cells \cite{Zhu2013}, photodetectors \cite{Aygar2016,Fang2017}, and on-chip sensors of multiple bio/chemical assays (multiplexers) \cite{Aouani2013,Aslan2016,Hsu2010,Lee2010a}.\\ 
\indent Here, we present the fabrication, optical characterization, and electromagnetic simulation of Au plasmonic structures inspired by the Sierpinski carpet (SC) deterministic fractal \cite{Mandelbrot1982}. Up to date, such SCs have been experimentally studied mainly in the far-infrared range \cite{Bao2007,Bao2008}, while at optical frequencies only calculations have been reported \cite{Volpe2011,Zhu2013}. We systematically investigated both experimentally and theoretically plasmonic SCs from the visible to mid-infrared (VIS-MIR) range. In particular, we realized for the first time a space-filling SC with five orders of complexity. We show that the fractal acts as a photonic quasi-crystal \cite{Janot1994}, due to the self-similar and hierarchical arrangement of its elements, with a multiband spectral response over the full investigated range, which can be tuned by controlling the fractal size, thickness, and order. We demonstrate that the resonances observed in the extinction spectra are diffraction-mediated surface plasmons. Furthermore, we illustrate by means of surface enhanced Raman spectroscopy (SERS) \cite{DeAngelis2011,Kneipp1997,Chirumamilla2014,Gopalakrishnan2014} that such plasmonic SCs are able to confine strong electromagnetic fields down to the nanoscale (sub-diffraction focusing) and on a broadband range of the electromagnetic spectrum, thereby providing a promising application for bio/chemical sensing.
\begin{figure*}[ht]
\centering
\includegraphics[width=0.9\textwidth]{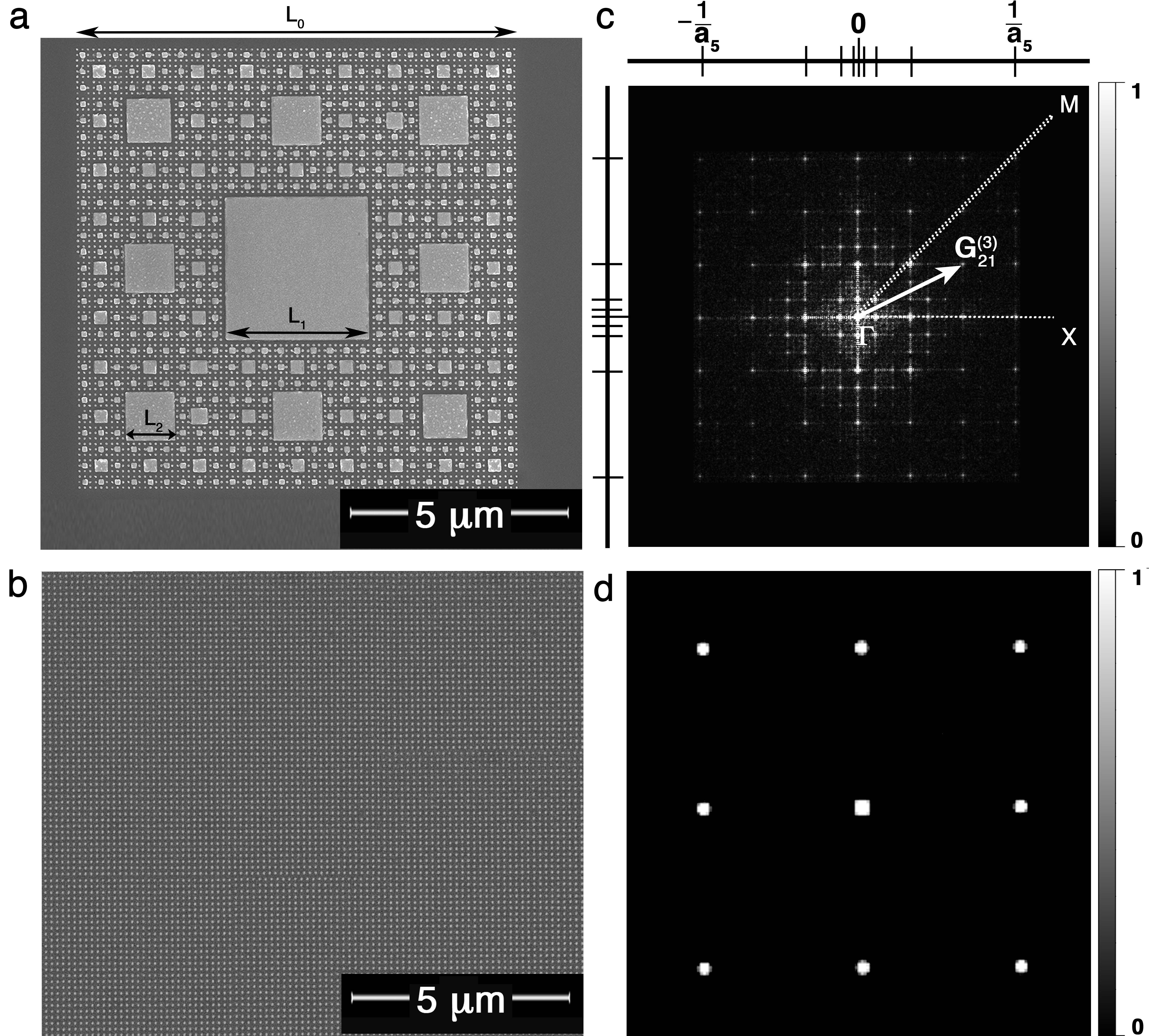}
\caption{Plasmonic Au SC. \textbf{a}, Scanning electron micrograph of a $35\pm3$ nm thick Au SC deposited on a Si/SiO$_{2}$ substrate for fractal order $t=5$. \textbf{b}, A $35\pm3$ nm thick Au periodic array with square size $L_{5}$, as a comparison. Fast Fourier transform of the SEM micrographs for the SC at $t=5$ (\textbf{c}) and the periodic array (\textbf{d}). The $\Delta=\overline{\Gamma X}$ and $\Sigma=\overline{\Gamma M}$ direction in the fractal reciprocal lattice are marked along with the first pseudo-Brillouin zones $[-\pi/a_{t},\pi/a_{t}]^{2}$ of the different orders.}
\label{fig:Figure1}
\end{figure*}
\section*{Results and discussion}
\label{sec:results}
\indent Deterministic fractals \cite{Mandelbrot1982} are self-similar objects generated by geometrical rules, having a non-integer (Hausdorff-Besicovitch) dimension. Sierpinski carpets can be generated by a recursive geometrical algorithm employing a Lindenmayer system implementation \cite{Lindenmayer1990} (see Supporting Information 1). Practically, our Au SCs are designed starting from a unit cell of side $L_{0}=10$ $\mu$m that is divided into a $3\times3$ array of sub-cells of lateral size $L_{1}=L_{0}3^{-1}$, with an Au square placed in the central sub-cell. By iteratively applying the same rule to the generated sub-cells of size $L_{t}=L_{0}3^{-t}$, it is possible to obtain fractals for higher orders $t$ of complexity. At each iteration the side of the sub-cells is reduced by a factor $\mathcal{L}=3$. Since the number of empty sub-cells in the SC increases by a factor $\mathcal{N}=8$ at every iteration, the fractal dimension is $d_{H}=\log{\mathcal{N}}/\log{\mathcal{L}}\approx1.89$, while the topological dimension of such a discrete system is zero.\\
\indent By employing electron-beam lithography, metal evaporation, and lift-off techniques (see Methods), we fabricated the first five orders of an Au SC on CaF$_{2}$ and Si/SiO$_{2}$ substrates (Supporting Information 2). The nominal thickness of the Au squares ranged from $25\pm3$ nm to $45\pm3$ nm, while their lateral sizes were $L_{1}=3.38\pm0.05$ $\mu$m, $L_{2}=1.12\pm0.01$ $\mu$m, $L_{3}=390\pm17$ nm, $L_{4}=130\pm7$ nm, and $L_{5}=44\pm3$ nm. The representative scanning electron microscopy (SEM) micrograph of the sample for $t=5$ in Figure \ref{fig:Figure1} show the excellent homogeneity and uniformity of the fractal structure. In addition, we realized periodic arrays of Au squares with lateral size $L_{t}$, as a reference. A periodic array with size $L_{5}$ is reported in Figure \ref{fig:Figure1}b.\\
\indent Our fractal structure can be regarded as a metallo-dielectric photonic crystal. However, the quasi-periodic \cite{Janot1994} and self-similar arrangement of the square elements forming the SC crystalline lattice breaks, due to its long-range order, the translational invariance property typical of periodic photonic crystals \cite{Sakoda2005}, while retaining a scale invariance. Therefore, fractals have also self-similar and non-discrete reciprocal lattices \cite{Negro2012}. In Figure \ref{fig:Figure1}c,d a comparison between the singular-continuous \cite{Negro2012} reciprocal lattice of the SC for $t=5$ and the discrete square reciprocal lattice of the periodic array with square size $L_{5}$ is provided, respectively, by computing the Fourier transform of their SEM micrographs. As confirmed by the box-counting method (see Supporting Information 3), the reciprocal lattice of the SC is self-similar. Also, the SC has a larger number of points in the reciprocal space than the periodic array, since its fractal reciprocal lattice is a superposition of five periodic lattices with different constants $a_{t}=3L_{t}$, under the periodic approximation \cite{Negro2012,Gambaudo2014}. Starting from the origin, each point of the SC reciprocal lattice in Figure \ref{fig:Figure1}c is defined by a reciprocal lattice vector $\left|\mathbf{G}_{ij}^{(t)}\right|=2\pi\sqrt{i^{2}+j^{2}}/a_{t}$. It is worth mentioning that also $\mathbf{G}_{ij}$ is scale-invariant as $\mathbf{G}_{ij}^{(t)}=3^{t-1}\mathbf{G}_{ij}^{(1)}$, thereby realizing a hierarchy of first pseudo-Brillouin zones $[-\pi/a_{t},\pi/a_{t}]^{2}$. It follows that the diffraction patterns and the related optical spectra of such fractals are themselves self-similar \cite{Allain1986,Janot1994,Negro2012}, with a larger number of allowed optical states than in periodic arrays (Figure \ref{fig:Figure1}d). All these features make fractals, such as the SC, an intriguing geometry for the realization of a novel class of photonic quasi-crystals \cite{Negro2012}.\\
\begin{figure*}[ht!]
\centering
\includegraphics[width=1\textwidth]{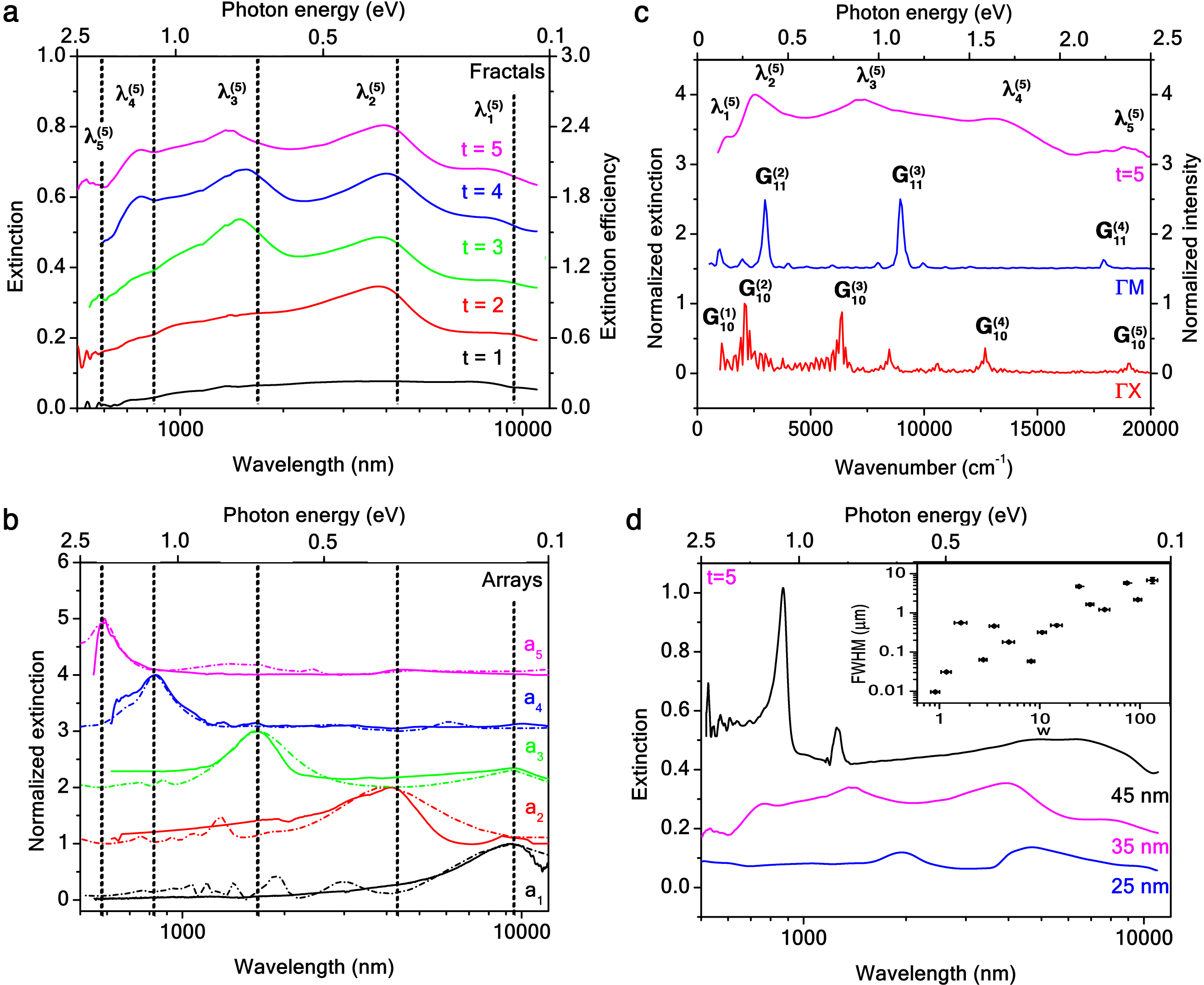}
\caption{Optical properties of plasmonic Au SCs. \textbf{a}, Experimental extinction and extinction efficiency spectra of $35\pm3$ nm thick SCs for $t=1$-5 orders. Extinction peaks $\lambda_{n}^{(5)}$ are marked. \textbf{b}, Experimental (solid curves) and calculated (dash-dot curves) normalized extinction of $35\pm3$ nm thick Au periodic arrays with lattice constants $a_{t}$. Dashed lines indicate the array LSPs. Curves in \textbf{a} and \textbf{b} are offset by 0.15. \textbf{c}, Experimental normalized extinction spectrum of the SC at $t=5$ (magenta solid curve) and normalized intensity of the points in the SC fast Fourier transform in Figure \ref{fig:Figure1}c, along the $\Delta=\overline{\Gamma X}$ (red solid curve) and $\Sigma=\overline{\Gamma M}$ (blue solid curve) directions. Curves are offset by 1.5. Note that the SC Fourier spectra were converted into photon energy units by $\hbar\omega=\hbar kc$. Extinction peaks $\lambda_{n}^{(5)}$ and reciprocal lattice vectors \textbf{G}$_{10}^{(t)}$ are marked. \textbf{d}, Experimental extinction spectra of SCs for $t=5$ with thicknesses $25\pm3$ nm (blue solid curve), $35\pm3$ nm (magenta solid curve), and $45\pm3$ nm (black solid curve). Inset, FWHM of the LSP modes as a function of the aspect ratio $w$ in log-log scale.}
\label{fig:Figure2}
\end{figure*}
\indent Surface plasmons are collective charge oscillations at the interface between a dielectric and a thin metal surface, which can be excited by light, for instance \cite{Kittel2005}. When the conductor size is much smaller than the wavelength of the incident light ($L<\lambda/10$), surface plasmons tend to be spatially localized around the structure \cite{Maier2007}.\\
\indent In the case of periodic and quasi-periodic photonic crystals, when the wavelength of a localized surface plasmon (LSP) resonance is slightly larger than the lattice constant, the photonic mode, due to the optical diffraction associated with the lattice, couples to the plasmonic mode of each particle to produce a hybrid mode \cite{Negro2012,Genet2007,Ghaemi1998,Ebbesen1998,Matsui2007,Bao2007,Bao2008,Gopinath2008,Kravets2008,Auguie2008} (Wood anomaly). In particular, when the wavelength of the incident light is larger than the lattice spacing (Rayleigh cutoff), all the diffracted waves other than the zeroth order are evanescent and all the particles are radiating in phase by dipolar coupling in the plane of the grating. This results from the momentum-matching condition \cite{Ghaemi1998,Ebbesen1998,Gopinath2008,Kravets2008} $\mathbf{q}=\mathbf{k}\sin{\theta}\pm\mathbf{g}_{ij}$, where $\mathbf{q}=\mathbf{k}\sqrt{\epsilon_{eff}}$ is the diffracted beam wavevector, $\epsilon_{eff}$ is the effective dielectric constant of the lattice, $\left|\mathbf{k}\right|=2\pi/\lambda$ is the wavevector of the incident light, $\theta$ is its angle of incidence with respect to the lattice surface normal, and $\mathbf{g}_{ij}$ are the lattice wavevectors. Herein, $\mathbf{g}_{ij}$ are the SC reciprocal vectors $\mathbf{G}_{ij}^{(t)}$. At normal incidence, diffraction-mediated LSPs can be excited when $\mathbf{k}^{(t)}\sqrt{\epsilon_{eff}}=\pm\mathbf{G}_{ij}^{(t)}$. We predict a set of four-fold degenerate LSP resonances, due to the square lattice symmetry, emerging in the Au SC optical spectra at self-similar wavelengths (critical modes \cite{Negro2012}) 
\begin{equation}
\lambda_{n}^{(t)}=\frac{a_{0}8^{-n/d_{H}}}{\sqrt{i^{2}+j^{2}}}\sqrt{\epsilon_{eff}},
\label{eq:lsp}
\end{equation}
where $n\in\left[1,t\right]$ is the $n$-th resonance at the $t$-th fractal order.\\
\begin{figure}[ht!]
\centering
\includegraphics[width=0.5\textwidth]{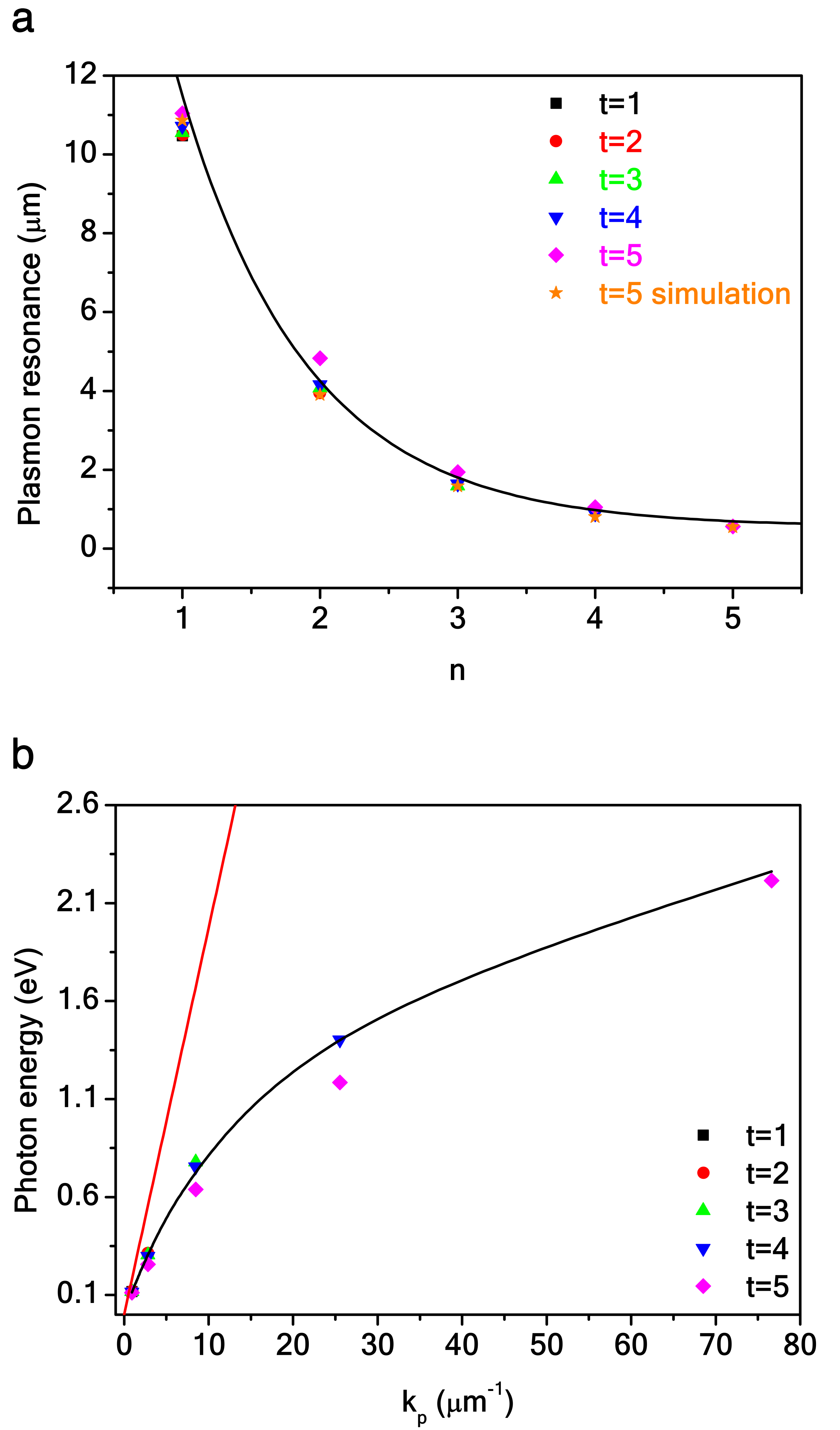}
\caption{Scaling of the Au SC plasmons. \textbf{a}, Experimental LSP resonances $\lambda_{n}^{(t)}$ as a function of their index $n$. The black solid line is the best fit of the experimental data given by Equation \ref{eq:lsp}, which is in agreement with electromagnetic simulations for $\lambda_{n}^{(5)}$ (orange star). The fit gives a fractal dimension value $d_{H}=1.92\pm0.04$. \textbf{b}, Experimental and calculated (black curve) LSP resonances as a function of LSP wavevector $k_{p}=\pi/L_{t}$. The red line represents the free-space dispersion of light $\hbar\omega=\hbar kc$. }
\label{fig:Figure3}
\end{figure}
\begin{figure*}[ht]
\centering
\includegraphics[width=0.93\textwidth]{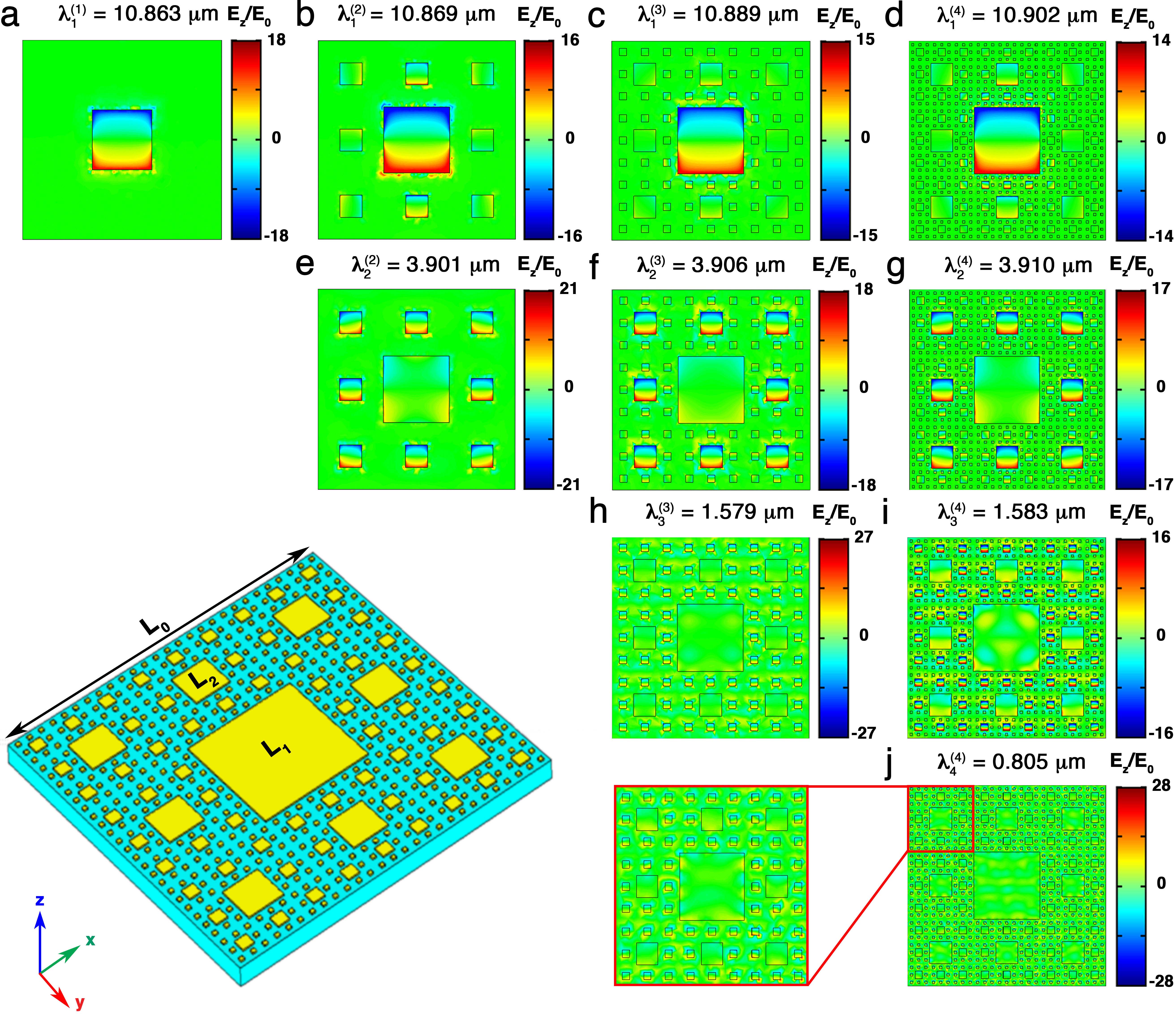}
\caption{Finite element method simulations of plasmonic Au SCs. Simulated electric near-field enhancement ($E_{z}/E_{0}$) distribution of SCs for orders $t=1-4$ (from left to right) at their resonances $\lambda_{1}^{(t)}$ (\textbf{a}-\textbf{d}), $\lambda_{2}^{(t)}$ (\textbf{e}-\textbf{g}), $\lambda_{3}^{(t)}$ (\textbf{h}-\textbf{i}), $\lambda_{4}^{(t)}$ (\textbf{j}). The incident electric field $E_{0}$ is polarized vertically and the phase is set to $\pi/4$ in order to maximize the field intensity. Each distribution is normalized to its maximum value. Inset, a sketch of the modeled Au SC deposited on a CaF$_{2}$ substrate.}
\label{fig:Figure4}
\end{figure*}
\indent In Figure \ref{fig:Figure2}a, we observed the zeroth-order (the incident and detected light are collinear) extinction $(1-T)$ spectra of the Au SCs exhibiting multiple resonances in the VIS-MIR range. The extinction efficiency (i.e., the extinction divided by the total area of the fractal) of the resonances is up to 185\%, thus showing an extraordinary extinction \cite{Lochbihler1994}. From the comparison with the extinction spectra of the periodic arrays (Figure \ref{fig:Figure2}b), it is possible to infer that the position of the SC resonances scales linearly with the lattice constant $a$ (Figure \ref{fig:Figure2}b). The SC resonances are blue-shifted with respect to those of the periodic arrays, due to the coupling between the squares belonging to the different fractal orders. As depicted in Figure \ref{fig:Figure2}c, the position of the resonances in the SC extinction correspond to the first reciprocal lattice vector $G_{10}^{(t)}$ of each fractal order. However, such modes occur at slightly smaller wavenumbers than the Fourier peaks in the $\overline{\Gamma X}$ and $\overline{\Gamma M}$ directions obtained from Figure \ref{fig:Figure1}c (i.e., at wavelengths slightly larger than the lattice constants $a_{t}$), thus confirming that the origin of such resonances in the Au SC extinction spectra is not due to a diffraction mechanism, but to diffraction-mediated LSPs.\\
\indent Furthermore, Au SCs are insensitive to the linear and diagonal polarization of the incident light, having a centrosymmetric geometry (see Supporting Information 4). This is an important feature in order to realize polarization-independent devices, since other fractal geometries \cite{Aslan2016,Gottheim2015,Miyamaru2008,Negro2012,Chirumamilla2017} and conventional plasmonic structures \cite{Genet2007,Maier2007} usually depend strongly on light polarization.\\
\begin{figure*}[ht]
\centering
\includegraphics[width=1\textwidth]{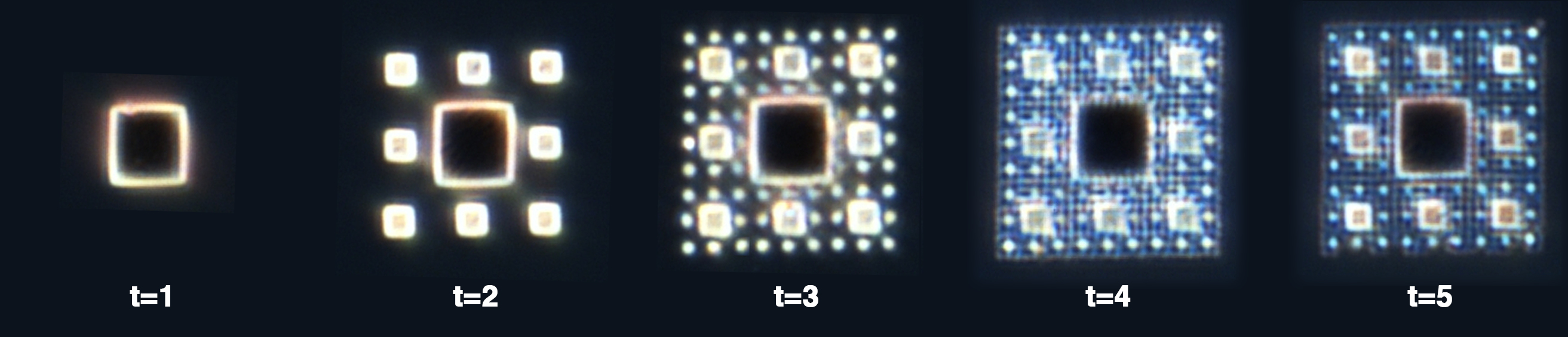}
\caption{Dark-field images of the SCs at different fractal orders $t=1$-5. Note that since the microscope halogen lamp is a white source, it does not emit at mid-infrared wavelengths. Thus, the first and second fractal orders diffract the incident light ($\lambda<<L_{t}$).}
\label{fig:Figure5}
\end{figure*}
\indent On the other hand, SCs do depend on the thickness of the metal squares, since they must be optically thick to achieve a large extinction. This implies that the thickness must be several times the skin depth of the metal \cite{Genet2007}. Typical skin depths are on the order of 30-170 nm for Au in the VIS-MIR range. As illustrated in Figure \ref{fig:Figure2}d for an Au SC at $t=5$, the extinction spectrum changes qualitatively for metal thicknesses ranging from $25\pm3$ nm to $45\pm3$ nm displaying sharp Fano resonances \cite{Lukyanchuk2010} in place of broad peaks (see Supporting Information 5). This suggests that the full-width-half-maximum (FWHM) of the LSPs depends on the aspect ratio (i.e., lateral size to thickness) $w_{t}=L_{t}/h$ \cite{Ebbesen1998}. Indeed, the inset in Figure \ref{fig:Figure2}d shows that the closer the aspect ratio to unity, the sharper and the shorter in wavelength the resonance (see Supporting Information 6). Therefore, the sharpest resonance occurs when the structures are isotropic (cubic), and the broadest when they are oblate.\\
The SC optical spectrum at a given fractal order $t$ exhibits $n=t$ resonances with wavelength $\lambda_{n}^{(t)}$ corresponding mainly to the first order of diffraction $(i,j)\equiv(1,0)$. Of such resonances, $n-1$ refer to the SC at order $t-1$, although shifted by a factor $\propto a_{t}^{-1}$ owing to the far-field diffraction coupling among the increased number of squares \cite{Maier2007}. In Figure \ref{fig:Figure3}a the wavelengths $\lambda_{n}^{(t)}$ of the peaks in the extinction measurement for each fractal order are plotted, along with the calculated data for $\lambda_{n}^{(5)}$. The best fit of the experimental data, which is in excellent agreement with the numerical simulations, is given by Equation \ref{eq:lsp} (where $d_{H}$ is treated as the fitting parameter), setting $(i,j)\equiv(1,0)$. The fit returns a fractal dimension $d_{H}=1.92\pm0.04$. As predicted, such LSP resonances are self-similar as their wavelength has the same scale-invariance law ($\mathcal{L}_{t}=8^{-t/d_{H}}$) as the SC, depending on the fractal dimension as an exponential law. From data in Figure \ref{fig:Figure2}a, we can draw in Figure \ref{fig:Figure3}b the dispersion relation of diffraction-mediated LSP modes. Here, $k_{p}=\pi/L_{t}=3\pi/a_{t}$ is the LSP wavevector of the SC, as shown in Figure \ref{fig:Figure4}. It follows that also $k_{p}$ is self-similar.\\
\indent In order to investigate the electromagnetic behavior of the plasmonic SCs, we computed their electromagnetic near-field spatial distributions by finite element method analysis. In Figure \ref{fig:Figure4}, the calculated electric near-field enhancement $E_{z}/E_{0}$ is depicted for the fractal orders $t=1$-4 at their LSP resonances. Our calculations show a resonant excitation of coupled dipolar antennas centered on the elements constituting the fractal. On the other hand, by increasing the fractal order, the system exhibits highly localized electromagnetic fields (hot spots) located on a sub-wavelength scale, resulting in additional plasmonic modes at shorter wavelengths. This mechanism provides a hierarchical multiscale of hot spots that transfer the excitations towards progressively smaller length scales, exhibiting large values of electric near-field enhancement. As a consequence of the SC fractal scaling the hot spot distribution of the resonant modes is self-similar. Two main phenomena can be distinguished. On one hand, a red-shift at a given LSP wavelength along with a decrease of its electric near-field intensity. As already stated for the far-field optical spectra, this is attributed to the coupling among structural elements of different size, with the difference that near-fields couple by dipolar interactions. On the other hand, new LSP modes arise at shorter wavelengths with higher intensity. Analogous considerations can be drawn for the magnetic field enhancement (see Supporting Information 7).\\
\begin{figure*}[hpb!]
\centering
\includegraphics[width=0.8\textwidth]{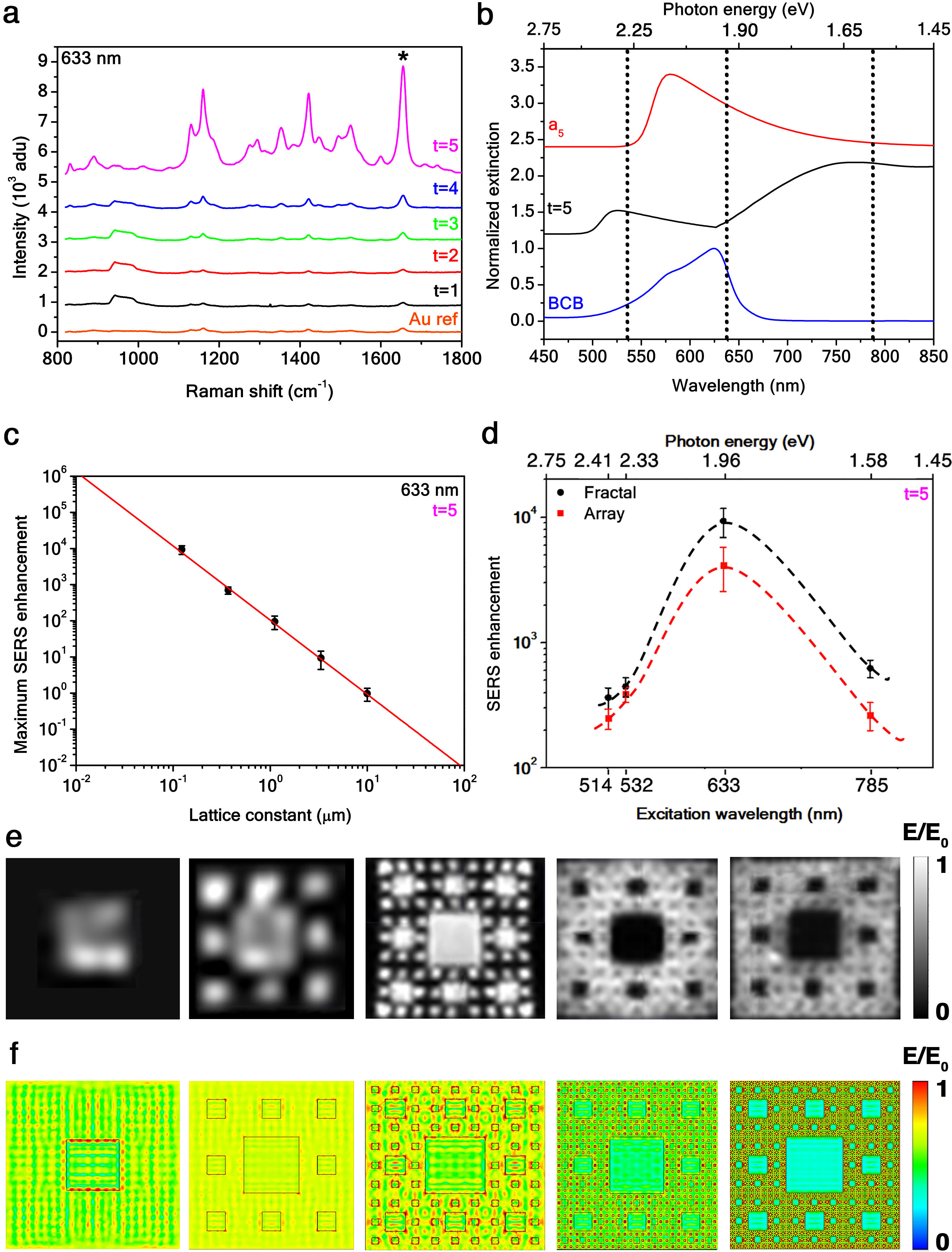}
\caption{Surface enhanced Raman spectroscopy on plasmonic Au SCs. \textbf{a}, Raman spectra at $\lambda_{ex}=633$ nm of BCB deposited on the SC for $t=1$-5 orders and on a reference Au film. Curves are offset by 0.20. The band at 800-1000 cm$^{-1}$ is due to the Si/SiO$_{2}$ substrate on which SCs were patterned. \textbf{b}, Normalized extinction spectra of BCB (blue solid curve), SC at $t=5$ (black solid curve), and periodic array at $L_{5}$ (red solid curve). Curves are offset by 0.20. Dotted lines represent Raman excitation wavelengths  $\lambda_{ex}$. \textbf{c}, SERS enhancement of the BCB vibrational mode $\omega^{\star}=1655$ $cm^{-1}$ at $\lambda_{ex}=633$ nm as a function of the SC lattice constant $a_{t}$, in log-log scale. \textbf{d}, SERS enhancement of the BCB vibrational mode $\omega^{\star}=1655$ $cm^{-1}$ as a function of $\lambda_{ex}$ for the SC at $t=5$ (black dots) and periodic array at $L_{5}$ (red squares). Dashed lines are guides for the eyes. Experimental electric field enhancement $E/E_{0}$ maps of $\omega^{\star}$ at $\lambda_{ex}=633$ nm for $t=1$-5 orders of the SC (\textbf{e}) and electromagnetic simulations of $E/E_{0}$ (\textbf{f}). Each map is normalized to its maximum value.}
\label{fig:Figure6}
\end{figure*}
\indent Another experimental evidence that LSPs can strongly couple by diffraction-mediated far-field interactions is presented in Figure \ref{fig:Figure5}. The micrographs, which refer to dark-field scattering maps of SCs illuminated using white light, illustrate characteristic localized spatial distributions with peculiar colors \cite{Gopinath2008}, which critically depend on the fractal order and on the distance between the squares. Hence, the interplay between short-range dipolar coupling (sub-wavelength near-field localization) and long-range far-field coupling (radiative multiple scattering) produces a variation of the near-field intensity and spatial distribution (Figure \ref{fig:Figure4}), and a wavelength shift of the LSP resonances along with a change in their intensity (Figure \ref{fig:Figure2}a) and far-field spatial distribution (Figure \ref{fig:Figure5}).\\
\indent In order to experimentally confirm the computational model, we investigated the SC electric near-field spatial distributions by SERS measurements on a thin layer ($12\pm2$ nm) of Brilliant Cresyl Blue (BCB) dye [(C$_{17}$H$_{20}$ClN$_{3}$O)$_{2}$ZnCl$_{2}$], as a probe deposited on the Au SCs. Average Raman spectra of the samples acquired at $\lambda_{ex}=633$ nm are plotted in Figure \ref{fig:Figure6}a for fractal orders $t=1$-5 and for an unpatterned Au film, as a reference. The excitation wavelength is resonant with the dye (Figure \ref{fig:Figure6}b). The Raman spectra show a series of vibrational bands, arising from the molecule, with an increasing intensity as a function of the fractal order. The enhancement of the Raman signal attributed to the SCs at higher orders is much stronger than that of the reference Au film, thus confirming the presence of hot spots. We selected the strongest BCB vibrational mode $\omega^{\star}=1655$ $cm^{-1}$, corresponding to the coupling of NH$_{2}$ scissor mode with the asymmetric stretch mode of C rings \cite{Nieckarz2013}, in order to evaluate the maximum SERS enhancement factor $EF_{sers}$ (see Supporting Information 8). The $EF_{sers}$ at $\lambda_{ex}=633$ nm as a function of the lattice constant $a_{t}$ of the fractals is reported in Figure \ref{fig:Figure6}c, while for the fractal at $t=5$ as a function of the incident wavelength is reported in Figure \ref{fig:Figure6}d. We found that $EF_{sers}$ increases with the fractal order as a power law. Despite BCB is resonant at $\lambda_{ex}=633$ nm, for orders $t=1$-3 the enhancement factor is very small as no LSP mode is in resonance (Figure \ref{fig:Figure6}b). For orders $t=4$-5, $EF_{sers}$ is highest as the excitation wavelength is not only resonant with the dye, but also with a LSP resonance of the SC. The resonant $EF_{sers}$ obtained at $\lambda_{ex}=633$ nm is about $10^{4}$, while in the non-resonant case, for instance at $\lambda_{ex}=785$ nm, is about $10^{3}$ (Figure \ref{fig:Figure6}d). It means that a maximum electric field enhancement factor of about $10$ is provided in the resonant case, while of about $5$ in the non-resonant case. In particular, the latter factor is not affected by the dye fluorescence, thus is purely a surface-enhanced electromagnetic effect. We infer that the plasmonic fractal has a broadband $EF_{sers}$, which is about twice higher in absolute value than an equivalent periodic array (Figure \ref{fig:Figure6}d), and such a structure could be used for multiplexing experiments with several assays absorbing in different ranges of the electromagnetic spectrum.\\
\indent In Figure \ref{fig:Figure6}e, the spatial distributions of electric field enhancement $E/E_{0}$, obtained from the the experimental Raman intensity maps of $\omega^{\star}$ at $\lambda_{ex}=633$ nm normalized to the signal of the reference Au film, are reported for fractal orders $t=1$-5. The contour plots present a maximum of contrast localized over the elements constituting the fractals for $t=1$-3 orders. Differently, for $t=4$-5 a large enhancement of the electric field is shown in between the structures of the previous orders, respectively in correspondence of the squares with size $L_{4}$ and $L_{5}$. Note that since the area of the laser spot is 0.785 $\mu$m$^{2}$, the instrument averages out over the structures $L_{4}$ and $L_{5}$. For instance, the intensity profile across the map at $t=5$ changes by a factor $\approx10$ from the central square $L_{1}$ to the surrounding smaller squares (see Supporting Information 8). Therefore, the smaller the element size, the larger the electric field localization occurring on it due to LSPs supported by the Au structures, which decay rapidly outside the squares. Our results are in good agreement with those obtained by electromagnetic simulations shown in Figure \ref{fig:Figure6}f, and with the work by Hsu et al. \cite{Hsu2010}. Notably, the computed map for $t=5$ has a maximum value of the electric field enhancement $E/E_{0}\approx13$ on $L_{5}$ squares at $\lambda_{ex}=633$ nm, which is comparable with the experimental value obtained $EF_{sers}^{1/4}\approx10$.\\
\indent In conclusion, we have demonstrated both experimentally and theoretically a novel multiband plasmonic antenna based on the compact design of a gold-based Sierpinski carpet fractal yielding multiresonant modes from the visible to mid-infrared range. This class of engineered devices offer promising applications for bio-chemical detection. As an example, we have carried out surface enhanced Raman spectroscopy on Brilliant Cresyl Blue molecules deposited onto a plasmonic Sierpinski carpet achieving a broadband enhancement factor up to $10^{4}$. The exploitation of such plasmonic fractal antennas to improve the versatility of Raman spectroscopy in the routine application to biology and chemistry.
\section*{Methods}
\label{sec:methods}
\subsection*{Sample fabrication}
\label{sec:fabrication}
\indent Sierpinski carpets were patterned by electron-beam lithography (Raith 150-Two) on CaF$_{2}$ and Si/SiO$_{2}$ substrates. A layer of approximately 160 nm of 950K poly-methyl-methacrylate (PMMA A2 1:1, 1\% in anisole) was deposited by spincoating on the substrates previously cleaned by O$_{2}$ plasma, then post-baked for 7 min on a hot plate at 180$^{\circ}$ C. In order to prevent charge accumulation due to the electron beam, a 10 nm thick Al conductive layer was deposited on CaF$_{2}$ substrates at a rate of 0.2 \AA/s by thermal evaporation (Kurt J. Lesker) at an operating pressure of $10^{-6}$ mbar. The electron beam was operated at 20 kV with a current of 35 pA. The SC patterns were exposed with a varying dose inversely proportional to the size of the squares of the fractal. The conductive layer was removed by washing it in KOH (1 M) for 10 seconds, followed by rinsing in deionized water. The resist was developed for 30 s in a cold (8$^{\circ}$ C) 1:3 mixture of MIBK:IPA. A $5/X$ nm (where $X$ is 25-45 nm) thick Ti/Au film was deposited on the substrates at a rate of 0.2 \AA/s by electron-beam evaporation (Kenosistec) at an operating pressure of $10^{-6}$ mbar. Lift-off was then performed by hot acetone. For Raman measurements, SCs were dip-coated for 1 h in 1 mM BCB aqueous solution, then rinsed in deionized water to wash the excess molecules in order to form a thin layer about 10 nm thick, and finally dried in nitrogen flow.
\subsection*{Sample characterization}
\label{sec:characterization}
\indent Scanning electron microscopy micrographs of SC samples deposited on Si/SiO$_{2}$ were acquired by FEI Helios NanoLab DualBeam 650. Atomic force microscopy height profiles of the samples deposited on Si/SiO$_{2}$ were measured in tapping mode by Bruker Innova in combination with V-type cantilever and SiN tips. Optical spectroscopy in transmission mode was performed with unpolarized and polarized light on SCs deposited on CaF$_{2}$ substrates, by Thermo Fisher FTIR spectrometer and Thermo Scientific Nicolet Continu$\mu$m microscope equipped with NO$_{2}$-cooled MCT and Si detectors, KBr and quartz beamsplitters, and a 15$\times$ (0.58 N.A.) Cassegrain objective. Surface enhanced Raman spectroscopy was carried out by Renishaw inVia micro-Raman microscope equipped with a 150$\times$ (0.95 N.A.) objective and $\lambda_{ex}=514$ nm, $\lambda_{ex}=532$ nm, $\lambda_{ex}=633$ nm, and $\lambda_{ex}=785$ nm laser sources at a power 0.5 mW and integration time 10 s. All spectra were calibrated with respect to the first-order silicon LO phonon peak at 520 cm$^{-1}$ and recorded in backscattering geometry at room temperature. Raman measurements were performed on BCB adsorbed on SCs and as a reference on a $35\pm3$ nm thick (3 nm root-mean-square roughness) Au film deposited on the same Si/SiO$_{2}$ substrate of the fractals. Raman maps of the BCB $\omega^{\star}=1655$ cm$^{-1}$ vibrational mode were scanned at 0.3 $\mu$m steps in both the directions in the plane of the sample. Renishaw WiRE 3.0 software was used to analyze the collected spectra, whose baseline was corrected to the third-order polynomial. Dark-field images of the SC samples were recorded by Nikon Eclipse upright microscope. Samples were illuminated with unpolarized white light by a 50 W halogen lamp in transmission mode. The light scattered by the SCs was collected with a 100$\times$ (0.96 N.A.) objective and imaged by a digital camera. 
\\\\
\begin{acknowledgement}
This work was supported by the European Union’s Horizon 2020 research and innovation programme under Grant Agreement No. 696656 Graphene Flagship -- Core1.\\
The authors declare no competing financial interests.
\end{acknowledgement}
\noindent \textbf{Supporting Information}. Detailed information on the generating algorithm of the fractals, additional experimental data, fractal analysis, optical response with polarized light, fitting of the extinction resonances, scaling of the plasmons, magnetic near-field simulations of the fractal, details on SERS measurements. 
\providecommand{\latin}[1]{#1}
\providecommand*\mcitethebibliography{\thebibliography}
\csname @ifundefined\endcsname{endmcitethebibliography}
  {\let\endmcitethebibliography\endthebibliography}{}

\end{document}